\documentclass[aps,prl,showpacs,twocolumn,superscriptaddress,nofootinbib]{revtex4-1}

\usepackage{graphicx}
\usepackage{color}      

\usepackage{bm}
\usepackage{amsmath}
\usepackage{times, txfonts}

\newcommand{\bra}[1]{\langle #1 \vert}
\newcommand{\ket}[1]{\vert #1 \rangle}

\newcommand{\mU}{\mathcal{U}}
\newcommand{\mV}{\mathcal{V}}

\newcommand{\nSigma}{\mathbf{\Sigma}}
\newcommand{\ii}{\textrm{i}}

\newcommand{\be}{\begin{equation}}
\newcommand{\ee}{\end{equation}}
\newcommand{\bea}{\begin{eqnarray}}
\newcommand{\eea}{\end{eqnarray}}

\newcommand{\nmax}{N_{\mbox{\footnotesize{max}}}}

\newcommand{\snn}{{\em S}$_{2n}$ }

\begin{document}

\title{Leading chiral three-nucleon forces along isotope chains in the calcium region}

\author{V. Som\`a}
\email{vittorio.soma@physik.tu-darmstadt.de}
\affiliation{Institut f\"ur Kernphysik, Technische Universit\"at Darmstadt, 64289 Darmstadt, Germany}
\affiliation{ExtreMe Matter Institute EMMI, GSI Helmholtzzentrum f\"ur Schwerionenforschung GmbH, 64291 Darmstadt, Germany}

\author{A. Cipollone}
\affiliation{Department of Physics, University of Surrey, Guildford GU2 7XH, UK}

\author{C. Barbieri}
\email{C.Barbieri@surrey.ac.uk}
\affiliation{Department of Physics, University of Surrey, Guildford GU2 7XH, UK}

\author{P. Navr\'atil}
\affiliation{TRIUMF, 4004 Westbrook Mall, Vancouver, BC, V6T 2A3, Canada}

\author{T. Duguet}
\email{thomas.duguet@cea.fr}
\affiliation{CEA-Saclay, IRFU/Service de Physique Nucl\'eaire, 91191 Gif-sur-Yvette, France}
\affiliation{National Superconducting Cyclotron Laboratory and Department of Physics and Astronomy,
Michigan State University, East Lansing, MI 48824, USA}

\date{\today}

\begin{abstract}
Three-nucleon forces (3NFs), and in particular terms of the Fujita-Miyazawa type, strongly influence the structure of neutron-rich exotic isotopes. {\em Ab-initio}  calculations have shown that chiral two- and three-nucleon interactions correctly reproduce binding energy systematics and neutron driplines of oxygen and nearby  isotopes.
Exploiting the novel self-consistent Gorkov-Green's function approach, we present the first investigation of Ar, K, Ca, Sc and Ti isotopic chains. Leading chiral 3N interactions are mandatory to reproduce the trend of binding energies throughout these chains and to obtain a good description of two-neutron separation energies. At the same time, nuclei in this mass region are systematically overbound by about 40 MeV and the $N=20$ magic gap is significantly overestimated. We conclude that {\em ab-initio} many-body calculations of mid-mass isotopic chains challenge modern theories of elementary nuclear interactions.
 \end{abstract}
\pacs{21.60.De, 21.30.-x, 21.45.Ff, 27.40.+z}

\maketitle

{\it Introduction}. 
Many-body interactions involving more than two nucleons play an important role in nuclear physics. They arise naturally, due to the internal structure of the nucleon and are deemed to be necessary to explain saturation properties of nucleonic matter~\cite{Soma:2008nn,Hebeler:2011snm,Carbone2013snm}. 
In finite systems, three-nucleon forces (3NFs) provide key mechanisms that control the shell evolution in exotic isotopes. 
Studies based on interactions derived from chiral effective field theory (EFT)~\cite{Epelbaum:2008ga} showed that leading two-pion 3NF terms (of the Fujita-Miyazawa type) induce changes in the location of traditional magic numbers and explain the anomalous position of the oxygen neutron dripline compared to neighbouring elements~\cite{Otsuka2010prl,Hergert2013prl,Cipollone2013prl}. 
As the effects of 3NFs become more evident in unstable neutron-rich isotopes, data from radioactive beam facilities worldwide  are fundamental to constrain theories of nuclear forces. In turn, progress in our understanding of nuclear interactions and accurate {\em ab-initio} calculations are crucial to achieve realistic predictions for unstable nuclei. 
This can eventually improve our insight into properties of isotopes that influence astrophysical processes but are not reachable experimentally.

Microscopic shell model calculations based on chiral 3NFs have been performed for isotopes up to the Ca chain~\cite{Otsuka2010prl,Holt2011jpg}.
Yet, full-fledged {\em ab-initio}  approaches have been so far limited to the oxygen mass region~\cite{Hergert2013prl,Cipollone2013prl,Lahde:2013uqa} plus a few heavier closed-shell systems~\cite{Hergert2013a,Binder:2012mk,Binder2013cc3nf}. In particular, Ref.~\cite{Cipollone2013prl} showed that next-to-next-to-leading order (NNLO) chiral 3NFs correctly reproduce the binding energies and the dripline location not only of oxygen but also of neighbouring nitrogen and fluorine isotopic chains.
This Letter extends {\em ab-initio} calculations and tests the performance of chiral interactions for masses around $Z=20$ where open-shell isotopes start to be dominant.

{\em Ab-initio} many-body methods capable of targeting the calcium region include self-consistent Green's function (SCGF)~\cite{Barbieri2009ni,Cipollone2013prl}, coupled cluster (CC)~\cite{Hagen:2012fb, Roth2012prl} and in-medium similarity renormalisation group (IM-SRG)~\cite{Tsukiyama2011prl,Hergert2013a} theories.
Such approaches make use of sophisticated and accurate many-body schemes but have been limited until very recently to the vicinity of doubly-magic systems. 
The standard formulation of SCGF, based on the Dyson equation, features e.g. the third order algebraic diagrammatic construction [ADC(3)], 
which includes all third order diagrams in the self-energy and resums higher-order ones non-perturbatively~\cite{Schirmer1983,Barbieri:2007Atoms}.
Recently, we have introduced a new method based on the Gorkov reformulation of SCGF formalism~\cite{Soma:2011GkvI} and produced proof-of-principle calculations~\cite{Soma:2013rc,Soma:2013GkvII} up to $^{74}$Ni. 
Presently implemented at second order, the Gorkov-GF approach can not only provide many observables associated with even-even nuclear ground states but can also access ground- and excited-state energies of odd-even neighbours. As such, it extends the reach of {\em ab-initio} calculations to a very significant portion of the nuclear chart.
This Letter reports on the first application of Gorkov-GF method to study the effect of chiral 3N interactions. Our focus is on absolute binding energies and two-neutron separation energies whose trend in calcium and neighbouring isotopes is only reproduced when leading 3NFs are incorporated. On the other hand, employed chiral forces overbind nuclei in the Ca region in contrast to what was seen in the oxygen region~\cite{Cipollone2013prl,Hergert2013prl}.

{\it Formalism}.
We start from the intrinsic Hamiltonian \hbox{$\hat{H}_{int}= \hat{T}-\hat{T}_{cm}+\hat{V}+\hat{W}$}, with the kinetic energy of the center of mass subtracted and $\hat{V}$ and $\hat{W}$ the two-nucleon (NN) and 3N interactions.  The Gorkov formalism breaks particle-number symmetry and targets the ground state, $\ket{\Psi_0}$, of the grand canonical Hamiltonian $\hat{\Omega}_{int}= \hat{H}_{int} - \mu_p \hat{Z} - \mu_n \hat{N}$ under the constraint that the correct particle number $A=N+Z$ is recovered on average: $Z=\bra{\Psi_0}\hat{Z}\ket{\Psi_0}$ and $N=\bra{\Psi_0}\hat{N}\ket{\Psi_0}$.
The central quantities of Gorkov-GF formalism are the normal and anomalous propagators associated with $\ket{\Psi_0}$
\begin{subequations}
\label{eq:Gab}
\begin{eqnarray}
\label{eq:Gab11}
G^{11}_{\alpha \beta} (\omega) &=&  \sum_{k} \left\{
\frac{\mU_{\alpha}^{k} \,\mU_{\beta}^{k*}}
{\omega-\omega_{k} + \ii \eta}
+ \frac{\bar{\mV}_{\alpha}^{k*} \, {\bar{\mV}_{\beta}^{k}}}{\omega+\omega_{k} - \ii \eta} \right\} \: ,\\
\label{eq:Gab12}
G^{12}_{\alpha \beta} (\omega) &=&   \sum_{k}
\left\{
\frac{\mU_{\alpha}^{k} \,\mV_{\beta}^{k*}}
{\omega-\omega_{k} + \ii \eta} + \frac{\bar{\mV}_{\alpha}^{k*} \, {\bar{\mU}_{\beta}^{k}}}{\omega+\omega_{k} - \ii \eta}
\right\}  \, ,\\
\label{eq:Gab21}
G^{21}_{\alpha \beta} (\omega) &=&  \sum_{k}
\left\{
\frac{\mV_{\alpha}^{k} \,\mU_{\beta}^{k*}}
{\omega-\omega_{k}  + \ii \eta}
+ \frac{\bar{\mU}_{\alpha}^{k*} \, {\bar{\mV}_{\beta}^{k}}}{\omega+\omega_{k} - \ii \eta}
\right\} \, ,\\
\label{eq:Gab22}
G^{22}_{\alpha \beta} (\omega) &=&   \sum_{k} \left\{
\frac{\mV_{\alpha}^{k} \,\mV_{\beta}^{k*}}
{\omega-\omega_{k} + \ii \eta}
+  \frac{\bar{\mU}_{\alpha}^{k*} \, {\bar{\mU}_{\beta}^{k}}}{\omega+\omega_{k} - \ii \eta} \right\} \: .
\end{eqnarray}
\end{subequations}
where greek indices $\alpha$,\,$\beta$, ...  label a complete orthonormal single-particle basis and barred quantities refer to time-reversed states~\cite{Soma:2011GkvI}.
The poles of the propagators are given by \hbox{$\omega_{k} \equiv \Omega_k - \Omega_0$}, where the index $k$ refers to normalized eigenstates of $\hat{\Omega}_{int}$ that fulfill
$\hat{\Omega}_{int} \, | \Psi_{k} \rangle = \Omega_{k} \, | \Psi_{k} \rangle$.
The residues of pole $\omega_{k}$ relate to the probability amplitudes 
$\mU^k$ ($\mV^k$) to reach state $| \Psi_{k} \rangle$ by adding (removing) a nucleon to (from) $| \Psi_{0} \rangle$.

Normal and anomalous one-body density matrices are
\begin{subequations}
\label{eq:allbdm}
\begin{equation}
\label{eq:nobdm}
\rho_{\alpha \beta} \equiv \langle \Psi_0 | a_\beta^{\dagger} a_\alpha | \Psi_0 \rangle  
= \sum_{k} \bar{\mV}_{\beta}^k \, \bar{\mV}_{\alpha}^{k*} \: , 
\end{equation}
\begin{equation}
\label{eq:aobdm}
\tilde{\rho}_{\alpha \beta} \equiv \langle \Psi_0 | \bar{a}_\beta a_\alpha | \Psi_0 \rangle
= \sum_{k} {\bar{\mU}_{\beta}^k} \, \bar{\mV}_{\alpha}^{k*}   \: ,
\end{equation}
\end{subequations}
whereas the binding energy is computed through the Koltun sum rule corrected for the presence of 3NFs~\cite{Carbone2013tnf,Cipollone2013prl},
\begin{equation}
\label{eq:koltun}
E^{\text{A}}_0=
 \sum_{\alpha \beta}
 \int_{C \uparrow} \frac{d \omega}{4 \pi i}  \,  G_{\alpha \beta}^{11} (\omega) \left[ T_{\beta \alpha}
+ \delta_{\beta \alpha} \left(\mu + \omega \right)  \right]   \;
 - \frac12  \bra{\Psi_0 } \hat{W}  \ket{\Psi_0 } \, .
\end{equation}

The  propagators~\eqref{eq:Gab} are solutions of the Gorkov equation,
\begin{equation}
\label{eq:gkv}
\left.
\left(
\begin{tabular}{c}
\hspace{-0.2cm} $T  + \Sigma^{11}(\omega)- \mu_k   \qquad \quad  \Sigma^{12}(\omega)$ \\
$\Sigma^{21}(\omega) \qquad \quad  \hspace{-0.4cm}  -T + \Sigma^{22}(\omega) + \mu_k $
\end{tabular}
\right)
\right|_{\omega_k}
\hspace{-0.1cm}
\left(
  \begin{array}{c}
\hspace{-0.1cm} \mU^k   \\
\hspace{-0.1cm} \mV^k
  \end{array} \hspace{-0.1cm} \right)
  \hspace{-0.1cm}
= \omega_{k}
\left(
  \begin{array}{c}
\hspace{-0.1cm} \mU^k   \\
\hspace{-0.1cm} \mV^k 
  \end{array} \hspace{-0.1cm} \right)  ,
\end{equation}
where the self-energy splits into static (first order) and dynamic contributions, $\nSigma(\omega)=\nSigma^{(\infty)}+\nSigma^{(dyn)}(\omega)$.
The inclusion of 3NFs in Dyson SCGF formalism is discussed in depth in~\cite{Carbone2013tnf}, with first applications in Refs.~\cite{Soma:2008nn,Cipollone2013prl,Carbone2013tnf}.
Here, we extend Gorkov GF approach~\cite{Soma:2011GkvI,Soma:2013GkvII} to 3NFs by following the prescription of Ref.~\cite{Cipollone2013prl}. The second order self-energy contains only interaction-irreducible diagrams and it is calculated using only an effective NN interaction, which includes contributions from $\hat{W}$.
However, the static self-energy acquires extra terms from interaction-reducible diagrams involving 3NFs.  Hence, we calculate $\nSigma^{(\infty)}$ as usual in terms the
NN interaction, $\hat{V}$, and then add the following 3NFs corrections:
\begin{subequations}
\label{eq:sig1_3nf}
\begin{eqnarray}
\Delta \Sigma_{\alpha \beta}^{11,(\infty)} &=&  - \left[\Delta \Sigma_{\bar\alpha \bar\beta}^{22,(\infty)} \right]^* =
 \frac12   \sum_{ \gamma \, \delta \,  \mu \, \nu }
 W_{\alpha \gamma \delta, \beta \mu \nu}  \, \rho_{\mu \gamma} \, \rho_{\nu \delta} \; ,
\\
\Delta \Sigma_{\alpha \beta}^{12,(\infty)} &=& \phantom{-} \left[\Delta \Sigma_{\beta \alpha}^{21,(\infty)} \right]^* = 
 \frac12   \sum_{ \gamma \, \delta \,  \mu \, \nu }
 W_{\alpha \bar\beta \delta, \mu \bar\nu \gamma}  \,  \tilde\rho_{\mu \nu}  \, \rho_{\gamma \delta}  \; .
\end{eqnarray}
\end{subequations}

Our calculations follow the sc0 approximation that is
extensively discussed in Ref.~\cite{Soma:2013GkvII} and start by solving
the Hatree-Fock Bogolioubov equation within
the chosen single-particle model space, including 3NFs in full. This provides the reference state
and the corresponding density matrices that are used to generate the effective NN interaction
according to Refs.~\cite{Carbone2013tnf,Cipollone2013prl}.
Following the sc0 prescription, $\nSigma^{(dyn)}$ remains unchanged throughout the rest of
the calculation while $\nSigma^{(\infty)}$ is evaluated in a self-consistent fashion in terms of
correlated density matrices, Eqs.~\eqref{eq:allbdm}. This is obtained through an iterative solution
of Eqs.~\eqref{eq:gkv}, \eqref{eq:allbdm} and~\eqref{eq:sig1_3nf}. This approach is the optimal
compromise between accuracy and computational cost and was found to be accurate to 1\% or better~\cite{Soma:2013GkvII}.

Once convergence is reached, the total energy is calculated according to Eq.~\eqref{eq:koltun}, where
the expectation value of $\hat{W}$ is obtained at first order in terms of the {\em correlated} normal density
matrix~\cite{Cipollone2013prl}
\begin{equation}
\bra{\Psi_0} \hat{W}  \ket{\Psi_0} \approx
 \frac16 
 \sum_{\alpha \, \beta \, \gamma \,  \mu \, \nu \, \xi}
 W_{\alpha \beta \gamma ,  \mu \nu \xi} \; \rho_{\mu \alpha} \, \rho_{\nu \beta} \, \rho_{\xi \gamma} \; .
\end{equation}
The contribution containing two anomalous density matrices was checked to be negligible and hence is not included here.

The present formalism assumes a $J^{\Pi}=0^+$ ground state and therefore targets even-even systems.
The ground state energy of odd-even neighbours is obtained through~\cite{Duguet:2001gs,Soma:2011GkvI}
\begin{equation}
  E^{A}_0 = \widetilde{E}^A + \omega_{k=0}  \; , \qquad  \mbox{for $A$ odd,} 
\end{equation}
where $\widetilde{E}^A$ is the energy of the odd-even nucleus computed as if it were an even-even
one, i.e. as a fully paired even number-parity state but forced to have an odd number
of particles on average, while $ \omega_{k=0}$ denotes the lowest pole energy extracted
from Eqs.~\eqref{eq:Gab} and~\eqref{eq:gkv} for that calculation.

{\it Results}.
Calculations were performed using chiral NN and 3N forces  evolved to low momentum scales through free-space similarity
renormalization group (SRG) techniques~\cite{Jurgenson2009prl}.
The original NN interaction is 
generated at next-to-next-to-next-to-leading order (N$^3$LO) with cutoff $\Lambda_{2N}$=500 MeV~\cite{Entem2003N3LO,Machleidt2011pr}, while a local N$^2$LO 3NF~\cite{Navratil2007tnf} with a reduced cutoff of $\Lambda_{3N}$=400~MeV  is employed. The 3NF low-energy constants $c_D$=-0.2 and $c_E$=0.098 are fitted to reproduce $^4$He binding energy~\cite{Roth2012prl}. The SRG evolution on the sole chiral NN interaction already generates 3N operators in the Hamiltonian, which we refer hereafter to as the ``induced'' 3NF. When the pre-existing chiral 3N interaction, including the two-pion exchange Fujita-Miyazawa contribution, is included, we refer to the ``full'' 3NF.
Calculations were performed in model spaces up to 14 harmonic oscillator (HO) shells [$N_{max} \equiv$~max~$(2n+l)$ = 13], including all NN matrix elements and limiting 3NF ones to configurations with $N_1$+$N_2$+$N_3\leq N^{3NF}_{max}$=16. An SRG cutoff $\lambda$=2.0~fm$^{-1}$ and HO frequency $\hbar\Omega$=28~MeV were used.

\begin{figure}[t]
\includegraphics[width=0.95\columnwidth,clip=true]{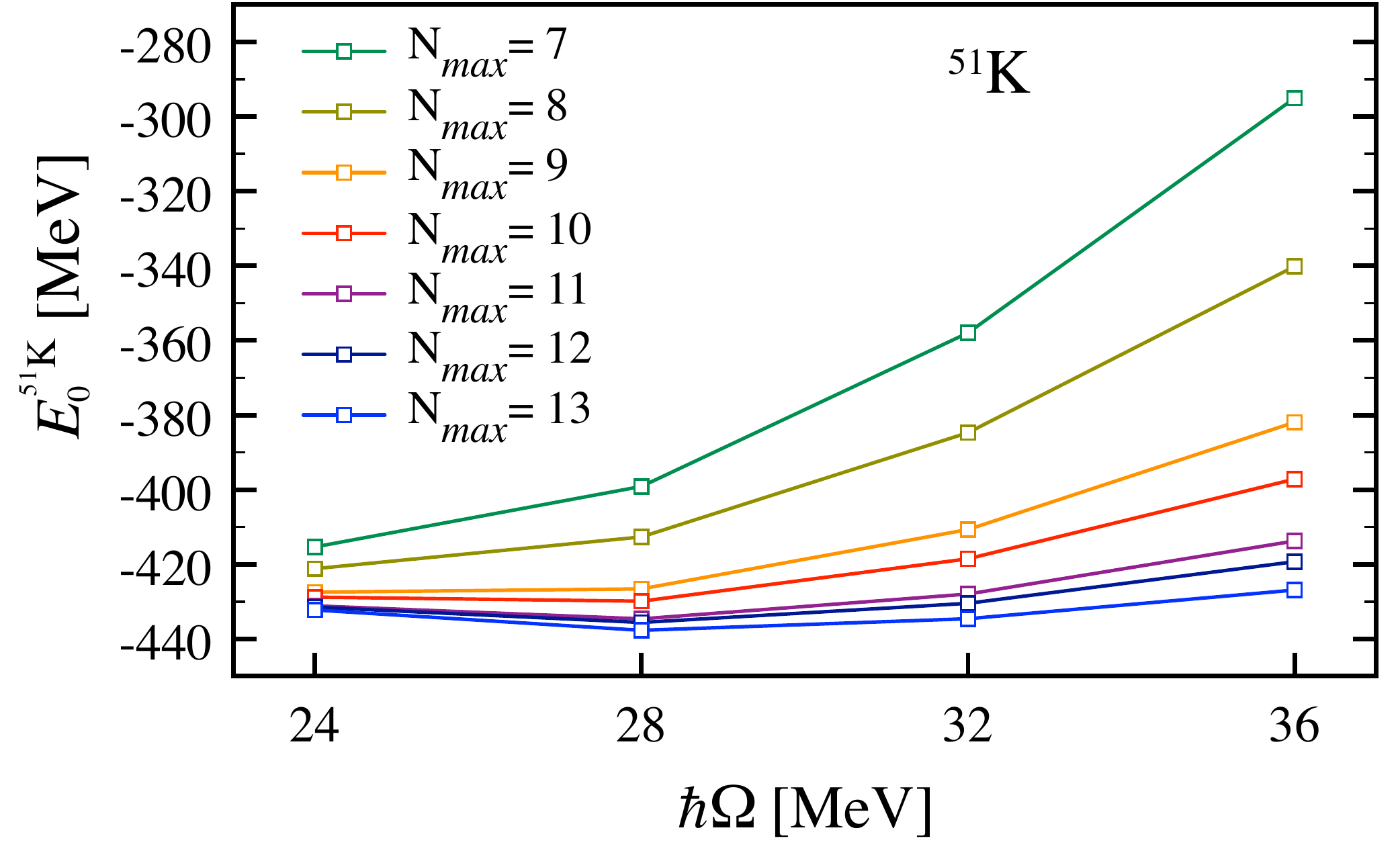}
\caption{(Color online) 
Convergence of the binging energy of $^{51}$K with respect to the basis size and HO frequency, for the full Hamiltonian.
}
\label{fig:MSconv}
\end{figure}

Figure \ref{fig:MSconv} shows the $^{51}$K binding energy as a function of the model space size and the HO frequency used. Being one of the heaviest nuclei considered here, $^{51}$K is representative of the slowest convergence obtained in this work. Changing the model space from $N_{max}$=12 to 13 lowers its ground-state energy by 2.1~MeV, which corresponds to about 0.5\% of the total binding energy. This is much smaller than the uncertainties resulting from truncating the many-body expansion of the self-energy at second order (see below). Other isotopes have similar speeds of convergence. For example, the change for the same variation of the model space induces a change of 1.7~MeV for $^{49}$K which slowly decreases to about 1~MeV in $^{40}$Ca. Thus, one expects convergence errors to cancel to a large extent when calculating two-neutron separation energies {\em S}$_{2n} \equiv E^{A}_0-E^{A-2}_0$, where the change in A is for the removal of two neutrons.
To test this we performed  exponential extrapolations of the calculated binding energies of a few nuclei, using the last few odd values of $N_{max}$. We found variations of at most $\approx$500~keV with respect to the  value calculated at $N_{max}$=13. Hence, we take this as an estimate of  the convergence error on computed {\em S}$_{2n}$.  In the following we present our results calculated for $N_{max}$=13 and $\hbar\Omega$=28~MeV, which corresponds to the minimum of the curve in Fig~\ref{fig:MSconv}. For isotopes beyond $N$=32, appropriate extrapolations and larger model spaces are required and will be considered in future works.

\begin{figure}[t]
\includegraphics[width=0.95\columnwidth,clip=true]{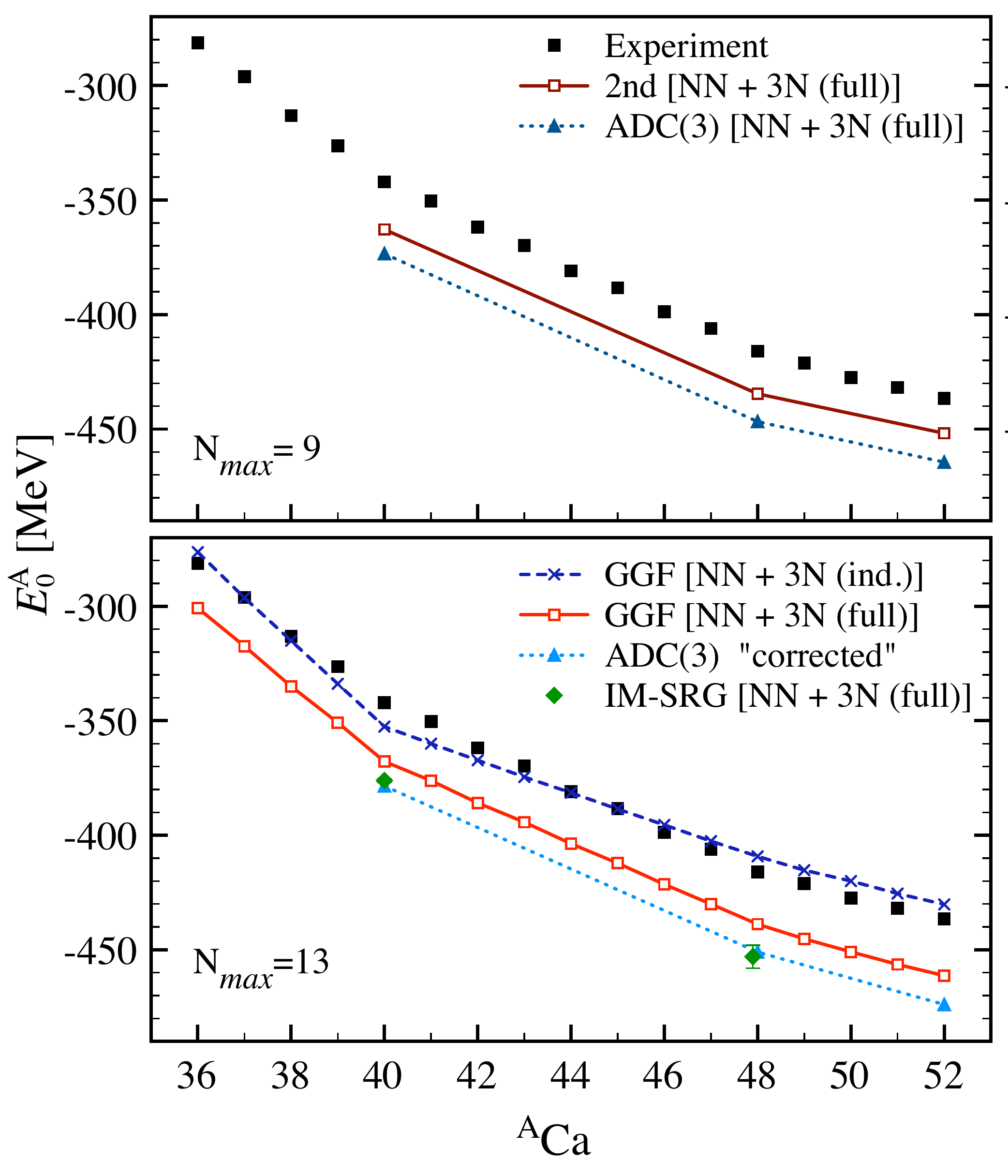}
\caption{(Color online) 
Experimental (full squares)~\cite{AME2012tables,Gallant2012prl,Wienholtz:2013nya} and calculated ground-state energies of Ca isotopes.
{\em Top panel}: second order Gorkov and Dyson-ADC(3) results for $^{40,48,52}$Ca obtained with a \hbox{$\nmax$ =9} model space and the full Hamiltonian.
{\em Bottom panel}: second order Gorkov results with NN plus induced (crosses) and NN plus full (open squares) 3NFs and \hbox{$\nmax$ =13}. 
Second order Gorkov results including full 3NF corrected for the ADC(3) correlation energy extracted from the top panel (dotted line with full triangles). IM-SRG results~\cite{Hergert2013a} are for the same 3NF and are extrapolated to infinite model space (diamonds with error bars).
}
\label{fig:BE_Ca}
\end{figure}

The accuracy of the many-body truncation of the self-energy at  second order must also be assessed. We do so by calculating closed-shell isotopes $^{40}$Ca, $^{48}$Ca and $^{52}$Ca for which Dyson ADC(3) calculations with $N_{max}$=9 can be performed and compared to second order Gorkov calculations (Gorkov-GF theory intrinsically reduces to Dyson-GF theory in closed-shell systems). Results in the top panel of Fig.~\ref{fig:BE_Ca} show that the correction from third- and higher-order diagrams is similar in the three isotopes. Specifically, we obtain $E_0^{ADC(3)-Dys}-E_0^{2nd-Gkv}=$~\hbox{-10.6}, -12.1  and  -12.6~MeV  that correspond to $\approx$2.7\% of the total binding energy. Assuming that these differences are converged with respect to the model space, we add them to our second order Gorkov results with $N_{max}$=13 and display the results in the bottom panel of  Fig.~\ref{fig:BE_Ca}. Resulting values agree well with IM-SRG calculations of $^{40}$Ca and $^{48}$Ca based on the same Hamiltonian~\cite{Hergert2013a}. This confirms the robustness of the present results across different many-body methods.
The error due to missing induced 4NFs was also estimated in Ref.~\cite{Hergert2013a} by varying the SRG cutoff over a (limited) range. Up to $\approx$1\% variations were found for masses $A\leq$~56 (e.g. less than 0.5\% for $^{40}$Ca and $^{48}$Ca) when changing $\lambda$ between 1.88 and 2.24 fm$^{-1}$. We take this estimate to be generally valid for all the present results.

A first important result of this Letter appears in the bottom panel of Fig.~\ref{fig:BE_Ca}, which compares the results obtained with NN plus induced and NN plus full 3NFs. The trend of the binding energy of Ca isotopes is predicted incorrectly by the induced 3NF alone. This is fully amended by the inclusion of leading chiral 3NFs. However, the latter introduce additional attraction that results in a systematic overbinding of ground-state energies throughout the whole chain. Analogous results are obtained for Ar, K, Sc and Ti  isotopic chains (not shown here), leading to the same conclusion regarding the role of the initial chiral 3NF in providing the correct trend and in generating overbinding at the same time. 
%

\begin{figure}[t]
\includegraphics[width=0.95\columnwidth,clip=true]{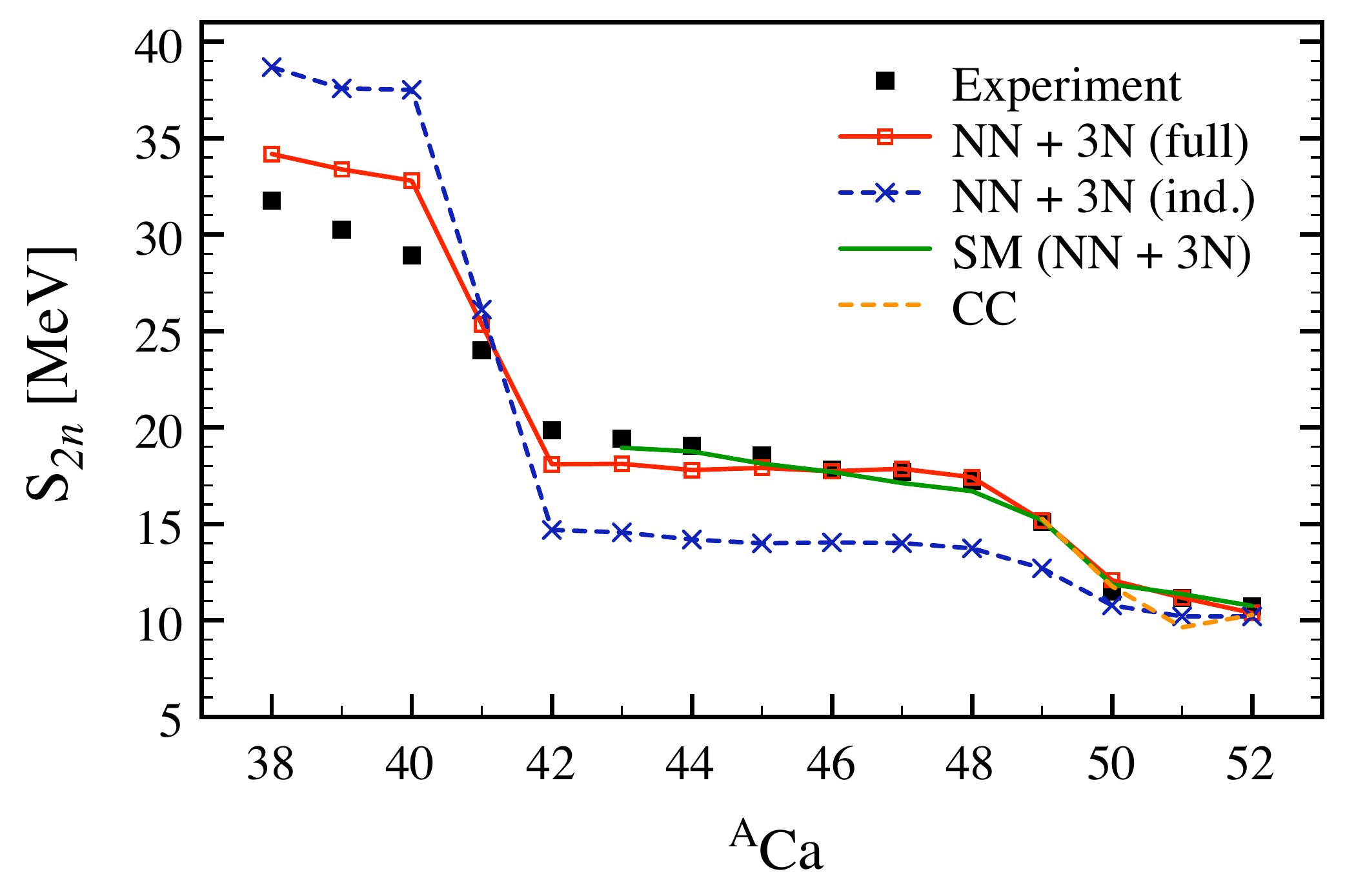}
\caption{(Color online) 
Two-nucleon separation energies, \snn, of Ca isotopes. Results for second order Gorkov calculation for are shown for the induced (crosses) and full (open squares) Hamiltonians and are compared to the experiment (full squares)~\cite{AME2012tables,Gallant2012prl,Wienholtz:2013nya}.
Results from shell model calculations with chiral 3NFs (full line)~\cite{Holt2011jpg,Wienholtz:2013nya} and coupled cluster (dashed line)~\cite{Hagen:2012fb} are also shown.
}
\label{fig:S2n_Ca}
\end{figure}

The NN plus induced 3N interaction, which originates from the NN-only N$^3$LO potential, generates a wrong slope in  Fig.~\ref{fig:BE_Ca} and exaggerates the kink at $^{40}$Ca.  The corresponding two-nucleon separation energies are shown in Fig.~\ref{fig:S2n_Ca}
and are significantly too large (small) for $N\leq$~20 ($N>$~20). Including chiral 3NFs correct this behaviour to a large extent and predict \snn  close to the experiment for isotopes above $^{42}$Ca. Figure~\ref{fig:S2n_Ca} also shows results for microscopic shell model~\cite{Holt2011jpg,Wienholtz:2013nya} and coupled cluster~\cite{Hagen:2012fb} calculations above  $^{41}$Ca and  $^{49}$Ca, respectively, which are based on similar chiral forces. Our calculations confirm and extend these results within a full-fledged {\em ab-initio}  approach for the first time. The results are quite remarkable, considering that NN+3N chiral interactions have been fitted solely to few-body data up to $A=4$.

The \snn jump between N=20 and N=22 is largely overestimated with the NN plus induced 3NFs, which confirms the findings of Refs.~\cite{Barbieri2009ni,Barbieri2009prl} based on the original NN interaction. The experimental Z=20 magic gap across $^{48}$Ca is
\hbox{$\Delta_\pi(^{48}{\rm Ca}) \equiv  2E_0^{{}^{48}{\rm Ca}} - E_0^{{}^{49}{\rm Sc}} -E_0^{{}^{47}{\rm K}}$}= 6.2~MeV, whereas it was found to be 10.5~MeV in Ref.~\cite{Barbieri2009prl}. The magic gap is somewhat larger in the present calculations, i.e. it is equal to 16.5~MeV with the NN plus induced 3NF and is reduced to 12.4~MeV including the full 3NF, which still overestimates experiment by about 6~MeV. 

Performing the integral in the Koltun sum rule~\eqref{eq:koltun} expresses the binding energy as a weighted sum of one-nucleon removal energies.
The systematic overbinding observed in the present results thus relates to a spectrum in the A-1 system (not shown here) that is too spread out. This has already been seen in Ref.~\cite{Barbieri2009ni} and is reflected in the excessive distance between major nuclear shells, or effective single-particle energies (ESPE)~\cite{Duguet2011,Soma:2011GkvI}.
In turn, the overestimated N=20 magic gap and the jump of the \snn between N=20 and N=22 relate to the exaggerated energy separation between {\em sd} and {\em pf} major shells generated by presently employed chiral interactions. Eventually, a too dilute ESPE spectrum translates into underestimated radii. 

Presently, ADC(3)-corrected energies with the NN plus full 3NF (Fig.~\ref{fig:BE_Ca}) overbind  $^{40}$Ca, $^{48}$Ca and $^{52}$Ca by 0.90, 0.73, and 0.72~MeV/A, respectively. It can be conjectured that such a behaviour correlates with a predicted saturation point of symmetric nuclear matter that is too bound and located at too high density compared to the empirical point. Recent calculations of homogeneous nuclear matter based on chiral interactions~\cite{Hebeler:2011snm,Carbone2013snm} predict a saturation point in the vicinity of the empirical point with an uncertainty that is compatible with the misplacement suggested by our analysis. However, such calculations use a different 3NF cutoff $\Lambda_{3N}$=500~MeV and different values of $c_D$ and $c_E$. 
Additional SCGF calculations as in Ref.~\cite{Carbone2013snm} but with the same NN+3N chiral interactions used here would help in confirming this conjecture.
 
\begin{figure}[t]
\includegraphics[width=0.95\columnwidth,clip=true]{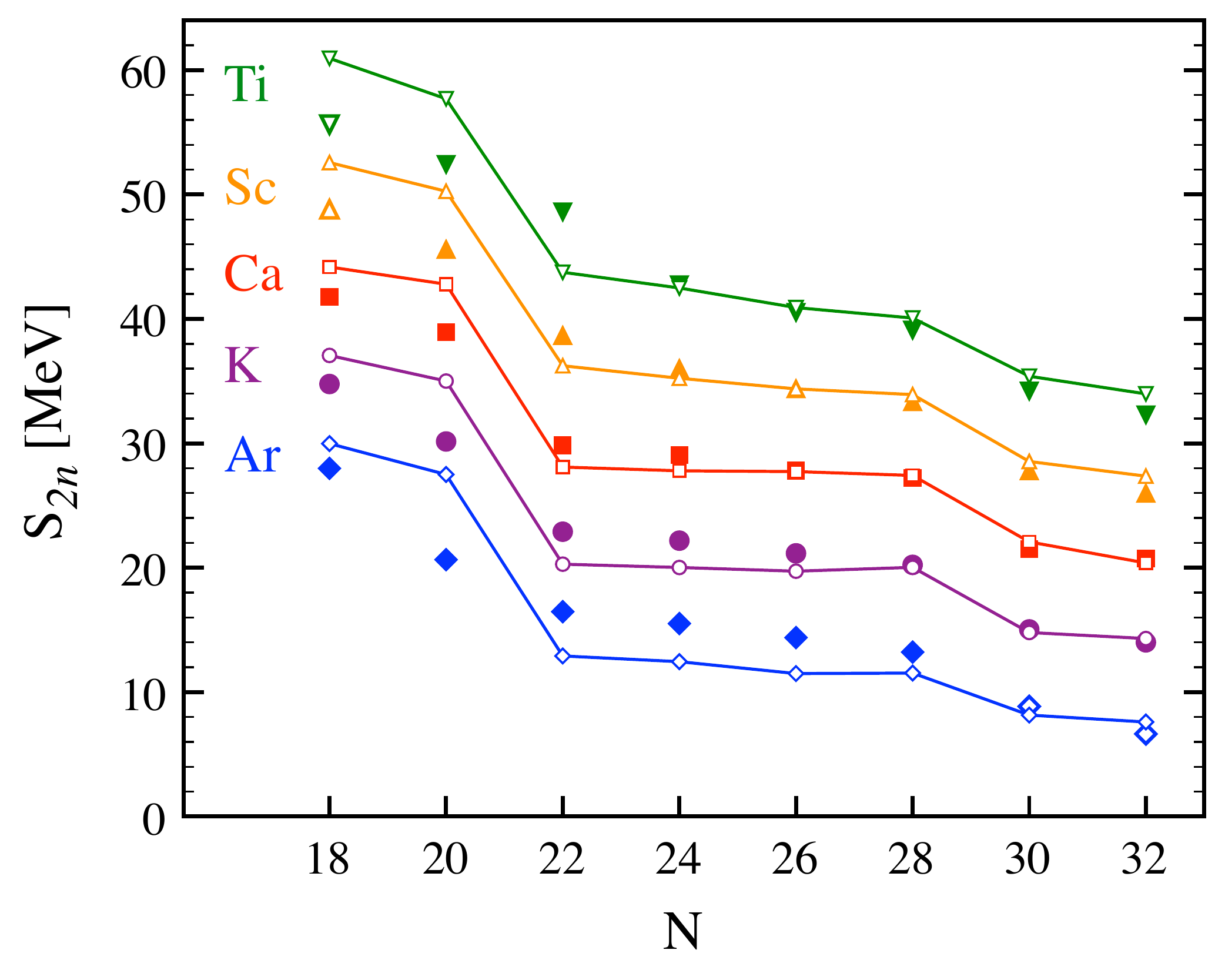}
\caption{(Color online) 
Two-neutron separation energies, \snn, along Ar, K, Ca, Sc and Ti isotopic chains. The experimental values (solid symbols)~\cite{AME2012tables,Gallant2012prl,Wienholtz:2013nya} are  compared to second order Gorkov calculations with the NN plus full 3NF (full lines).  Values for K, Ca, Sc and Ti are respectively shifted by +5~MeV, 10~MeV, 15~MeV and 20~MeV for display purposes. Isolated open symbols are AME2012 extrapolations of experimental data~\cite{AME2012tables}.
}
\label{fig:S2n_all}
\end{figure}

The systematic of \snn obtained with the NN plus full 3NF is displayed in Fig.~\ref{fig:S2n_all} along Ar, K, Ca, Sc and Ti isotopic chains, up to N=32. When the neutron chemical potential lies within the {\em pf} shell, predicted \snn reproduce experiment to good accuracy without adjusting any parameter beyond $A=4$ data. Still, the quality slightly deteriorates as the proton chemical potential moves down into the {\em sd} shell, i.e. going from Ca to K and Ar elements. The increasing underestimation of the \snn is consistent with a too large gap between proton {\em sd} and {\em pf} major shells that prevents quadrupole neutron-proton correlations to switch on. The too large jump of the \snn between N=20 and N=22 is visible for all elements and becomes particularly pronounced as one moves away from the proton magic $^{40}$Ca nucleus where the experimental jump is progressively washed out. At N=18, the situation deteriorates when going from $^{38}$Ca to $^{41}$Sc and $^{42}$Ti (but not going to $^{37}$K and $^{36}$Ar), i.e. when the proton chemical potential moves up into the {\em pf} shell. This is again consistent with an exaggerated  shell gap between {\em sd} {\em pf} shells that prevents neutron-proton correlations to switch on when the chemical potentials sit on both sides of the gap.

{\it Conclusions}.
We have reported on the first applications of second order Gorkov SCGF formalism with two- and three-body forces. The present approach allows for the systematic {\em ab-initio} description of mid-mass open-shell nuclei, including odd-even systems. This opens up the possibility to investigate large portions of the nuclear chart that were previously inaccessible to {\em ab-initio}  theory.  We have focused on the ability of leading chiral 3NFs to describe absolute binding and two-neutron separation energies along Ar, K, Ca, Sc and Ti chains, up to N=32. While available NN+3N chiral interactions typically perform well in the vicinity of oxygen isotopes, they were never tested in full for heavier masses. Leading 3NFs are found to be mandatory to reproduce the correct trend of binding energies for all isotopes, analogously to what was observed in lighter N, O and F chains. Overall, the systematic of two-neutron separation energies is reproduced with an impressive quality. Still, absolute binding energies are systematically overestimated throughout the Z$\approx$20 mass region and the magic character of N=20 and/or Z=20 nuclei is exaggerated by the employed NN+3N chiral forces. This can be traced back to the fact that the latter make deeply bound effective single-nucleon shells too spread out and the distance between major shells too pronounced. It is conjectured that these unwanted features relate to a saturation point of symmetric nuclear matter located at a slightly too large binding energy/density compared to the empirical point. Eventually, we conclude that {\em ab-initio} many-body calculations of mid-mass isotopic chains can now challenge modern theories of elementary nuclear interactions. While further improving the accuracy of cutting edge many-body methods, it will be of interest to investigate which particular features of available chiral interactions need to be improved on.

{\it Acknowledgements}. This work was supported by the DFG through Grant No. SFB 634, by the Helmholtz Alliance Program, Contract No. HA216/EMMI, by the United Kingdom Science and Technology Facilities Council (STFC) under Grant ST/J000051/1 and by the Natural Sciences and Engineering Research Council of Canada (NSERC), Grant No. 401945-2011. TRIUMF receives funding via a contribution through the National Research Council Canada. 
Calculations were performed using HPC resources from GENCI-CCRT (Grant No. 2013-050707) and the DiRAC Data Analytic system at the University of Cambridge (BIS National E-infrastructure capital grant No. ST/K001590/1 and STFC grants No. ST/H008861/1, ST/H00887X/1, and ST/K00333X/1).


\bibliography{Gorkov_3NF_Ca_chains}

\end{document}